\documentclass[9pt,twocolumn,twoside]{opticajnl} 
\journal{opticajournal} 
\setboolean{shortarticle}{true}

\usepackage{lineno} 
\usepackage{placeins}
\usepackage{booktabs}
\title{RF heating-enhanced photoacoustic tomography}

\author[1,2,3,*]{Skyler P. Selvin} 
\author[3]{XuanHao Wang} 
\author[3]{Handi Deng} 
\author[3]{Bohua Chen} 
\author[3,4,5]{Cheng Ma}

\affil[1]{Department of Electrical Engineering, Stanford University, Stanford, USA} 
\affil[2]{Global Engineering Internship Program, School of Engineering, Stanford University, Stanford, USA} 
\affil[3]{Tsinghua University, Department of Electronic Engineering, Beijing National Research Center for Information Science and Technology, Beijing 100084, China} 
\affil[4]{Institute for Precision Healthcare, Tsinghua University, Beijing 100084, China}
\affil[5]{IDG/McGovern Institute of Brain Research, Beijing 100084, China}
\affil[*]{selvin@stanford.edu}

\dates{}
\doi{Submitted version of paper published in Optics Letters: \href{https://doi.org/10.1364/OL.590278} {https://doi.org/10.1364/OL.590278}.}

\begin{abstract} Photoacoustic tomography (PAT) and thermoacoustic tomography (TAT) both leverage acoustic signals generated by electromagnetic absorption to noninvasively image deep tissues. PAT operates by detecting optical absorption, whereas TAT targets radiofrequency (RF) absorption, providing complementary information on tissue composition and structure. Combining these modalities into a single system promises richer contrast but remains difficult due to the expense and complexity of the RF source. Here, we show that PAT can be integrated with a low-cost RF heater and used to image both optical and RF absorption in tissue phantoms. RF Heating-Enhanced Photoacoustic Tomography (HEPAT) maps RF absorption via temperature-dependent changes in thermomechanical properties, which enables the use of slow, inexpensive RF subsystems and provides an additional layer of contrast. HEPAT therefore provides distinct, complementary contrast relative to existing photoacoustic imaging systems, expanding specificity and diagnostic power while opening new avenues for studying temperature-related tissue phenomena. \end{abstract}

\setboolean{displaycopyright}{false} 

\begin{document}

\maketitle


Photoacoustic tomography (PAT) is an imaging technique that has had a transformative impact on biomedical imaging \cite{wang2009multiscale}. PAT creates deep maps of optical absorption with ultrasound-level resolution (0.1 to 0.5 mm) and is excellent at visualizing vasculature due to hemoglobin's high optical absorption. However, many tissues do not demonstrate optical absorption contrast and therefore cannot be differentiated in conventional PAT. Attempts have been made to expand PAT's contrast \cite{fu2019photoacoustic}, including the use of exogenous contrast agents \cite{yang2009nanoparticles}, multiple optical wavelengths \cite{cox2009estimating, Shi2019}, and adding pulsed RF emitters to acquire RF absorption contrast \cite{pramanik2008breast,ku2005thermoacoustic,ke2012performance}. However, these methods are complex and expensive and still deliver limited contrast. It is therefore critical to continue developing systems that can expand PAT's contrast.

To this end, we look to the physics that generates a PA signal. Pulsed light incident on the sample surface is scattered and absorbed by structures $\sim$cm deep, causing them to thermally expand and emit ultrasound. These acoustic waves are then imaged with conventional ultrasound technology, and the resulting image maps the optical absorption deep within the sample. The signal amplitude $p_0$ is determined by how a given material produces acoustic waves from incident electromagnetic waves: $p_0 = \Gamma \eta \mu F$, where $\mu$ is the absorption coefficient, $F$ is the fluence, $\eta$ is the fraction of absorbed radiation that is converted to heat, and $\Gamma$, the Gr\"uneisen parameter, captures the thermomechanical properties of the material. PA contrast is governed primarily by the absorption coefficient $\mu$ and the Gr\"uneisen parameter $\Gamma$ (usually $\eta \approx 1$).

Tissues experience different electromagnetic absorption at different frequencies, and therefore there have been efforts to expand the range of illumination frequencies \cite{Shi2019,cox2009estimating,lazebnik2007large}. Thermoacoustic tomography (TAT) uses pulsed RF emissions in place of light, creating images mapping RF absorption \cite{pramanik2008breast,ku2005thermoacoustic,ke2012performance,lazebnik2007large,pramanik2009thermoacoustic}. Because RF absorption relies on different material resonances from optical absorption, TAT provides complementary contrast to PAT. The two have been combined to make dual-contrast systems \cite{pramanik2008breast,ku2005thermoacoustic,ke2012performance,pramanik2009thermoacoustic}. However, these systems are complex due to the intricacy of the pulsed RF source, and TAT's spatial resolution is often limited by the longer RF pulse width.

The Gr\"uneisen parameter $\Gamma$ is a thermomechanical material property. For many biological tissues, $\Gamma$ depends strongly on temperature \cite{Shi2019,Ma2016Grueneisen,shah2008photoacoustic,larina2005real,nikitin2012temperature,Wang2011ThermalIVPA,Liang2018GrueneisenLipids}. This effect has been used to monitor tissue temperature for thermotherapy \cite{shah2008photoacoustic,nikitin2012temperature}.
Fatty tissues experience a decrease in $\Gamma$ when heated, and watery tissues experience the opposite. Thus, measuring $d\Gamma/dT$ can aid in discriminating especially fatty or watery tissues such as plaques and tumors \cite{Wang2011ThermalIVPA, ku2005thermoacoustic, tian2015dualpulse}. 
In Gr\"uneisen-relaxation PAT, an initial heating pulse transiently raises the local temperature and modulates $\Gamma$, and subtracting the subsequent probe-pulse signals provides contrast related to $d\Gamma/dT$; in practice, however, the resulting contrast is often still dominated by optical absorption. \cite{Shi2019, Ma2016Grueneisen, tian2015dualpulse}

\begin{figure}[t]
    \centering
    \includegraphics[width=\linewidth]{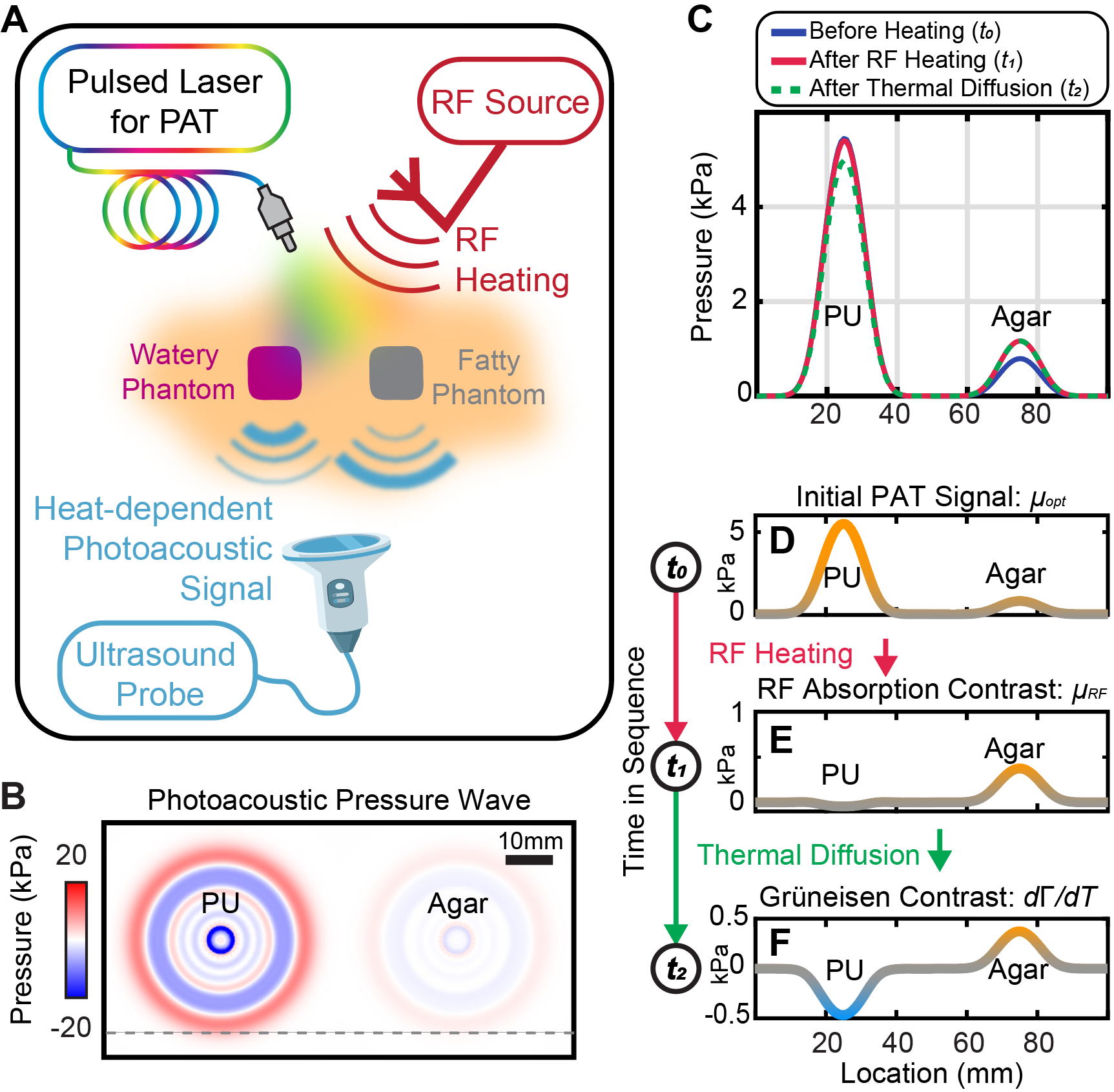}
    \caption{(A) Schematic of a HEPAT setup in which an RF source heats a sample and a pulsed laser generates a PA signal that is then detected with an ultrasound imaging system. As the sample's temperature rises, changes in its thermomechanical properties alter the PA signal, ultimately enabling additional contrasts. (B) Simulation showing a snapshot of the propagating photoacoustic pressure wave 15~$\mu$s after a single-pulse optical excitation in a tissue phantom containing a non-RF-absorbing region (polyurethane, PU) and a highly RF-absorbing region (agar). (C) Cross-sectional profiles along the dotted gray line in (B) before RF heating at $t_0$, immediately after heating at $t_1$, and after sufficient time for thermal diffusion at $t_2$. Immediately after heating, only the agar region shows a marked change in amplitude (demonstrating RF absorption contrast), whereas after thermal diffusion, both the agar and PU regions exhibit altered signals due to their temperature-dependent Gr\"uneisen parameters. This allows us to separate PA contrast, RF absorption contrast, and $d\Gamma/dT$ (Gr\"uneisen) contrast as shown in (D--F). (D) PA pressure before heating ($t_0$). (E) Pressure difference due to RF absorption but without thermal diffusion ($p(t_1) - p(t_0)$). (F) Pressure difference after thermal diffusion ($p(t_2) - p(t_0)$). Note that the opposite signs in $d\Gamma/dT$ contrast clearly show that the two targets are composed of different materials.} \label{fig:fig1}
\end{figure}

To simplify TAT's RF hardware, increase its resolution, and enable $d\Gamma/dT$ contrast, we integrate an RF heater into an existing PAT system to realize HEPAT, introducing both temperature-sensitive Gr\"uneisen parameter contrast and RF absorption contrast (Fig.~\ref{fig:fig1}). We acquire PAT images before and after RF heating, and by analyzing the heating-induced changes in the PAT signal, we map both the RF absorption coefficients and $d\Gamma/dT$ throughout the sample, thus demonstrating triple contrast: optical absorption $\mu_{opt}$, RF absorption $\mu_{RF}$, and thermomechanical $d\Gamma/dT$. Using $d\Gamma/dT$ itself as an imaging contrast uniquely enables emulating TAT functionality with much cheaper non-pulsed RF subsystems and introduces an additional contrast layer that can be isolated by leveraging thermal diffusion. Further, HEPAT's resolution is not limited by the pulse width of the RF emitter, as is often the case with TAT. 

To understand the contrasts of HEPAT, we analyze the effect of RF heating on PA signal intensity. Continuous RF radiation with intensity $I$ incident on a sample induces a local temperature change \( dT/dt = \mu_{RF} I / C_p \rho \), where $C_p$ is the isobaric specific heat capacity and $\rho$ is the mass density. RF in the high MHz to low GHz has a penetration depth in tissue that is comparable to that of PAT ($\sim$ cm) and can therefore heat the entire image volume \cite{gabriel1996dielectric,lazebnik2007large}. The PA signal amplitude changes when $\Gamma$ depends on temperature: \( \Gamma(T) = \beta c^2 / C_p \), where $\beta$ is the volumetric thermal expansion coefficient and $c$ is the speed of sound. For many materials and tissues, at least one of $\beta$, $C_p$, or $c$ changes with temperature \cite{shah2008photoacoustic,larina2005real,nikitin2012temperature,lou2010temperature,Wang2011ThermalIVPA}.

During RF exposure, areas with high $\mu_{RF}$ will increase in temperature. If $\Gamma$ is temperature-insensitive, heating will not modulate the PAT signal even when RF is absorbed and HEPAT will supply no additional contrast. However, usually $d\Gamma/dT \neq 0$ and both RF absorption and $d\Gamma/dT$ can be mapped by acquiring three PAT images at well-defined time points: (i) a baseline image at time $t_0$, before any RF heating; (ii) a “heated” image at time $t_1$, immediately at the end of heating, when the RF-absorbing region has been warmed but heat has not yet diffused into the RF-transparent regions; and (iii) a “diffused” image at time $t_2$, after heat diffuses into adjacent RF-transparent areas. 
The difference between the images at $t_1$ and $t_0$ highlights regions that absorb RF and therefore maps $\mu_{\mathrm{RF}}$, whereas the difference between the images at $t_2$ and $t_0$ primarily reflects $d\Gamma/dT$, since by $t_2$ the temperature rise has spread throughout the sample.  
Therefore, by taking successive PAT images at times before and after thermal diffusion, we can map $\mu_{RF}$ and $d\Gamma/dT$ separately.

Fig.~\ref{fig:fig1}B--F illustrates the HEPAT procedure using targets made of non-RF-absorbing polyurethane (PU) and highly RF-absorbing agarose gel (agar) that mimic fatty and watery tissues, respectively. Agar is mostly water and has high absorption at 2.45~GHz with $\mu_{RF} \approx 1~\mathrm{cm}^{-1}$, while polyurethane and fats have nearly no absorption \cite{gabriel1996dielectric}. The thermomechanical properties of agar and PU also differ: for agar, we expect about a 5\% increase in $\Gamma$ per $^\circ$C increase around room temperature, while $\Gamma$ will decrease by about 1\% per $^\circ$C in PU \cite{Liang2018GrueneisenLipids,larina2005real,nikitin2012temperature,Wang2011ThermalIVPA}.

We simulated a HEPAT experiment using a multiphysics model with coupled heat transfer and mechanical physics (Fig.~\ref{fig:fig1}B--F). Both targets started at room temperature (20~$^\circ$C). A single optical pulse generates an initial photoacoustic pressure distribution $p_0$, which then launches a propagating acoustic wavefield in the medium (Fig.~\ref{fig:fig1}B). Next, the agar was heated uniformly to 30~$^\circ$C with an RF heater, while the PU target, which did not absorb RF, remained at room temperature. A second PAT image then revealed signal changes exclusively in the agar phantom, demonstrating $\mu_{RF}$ contrast (Fig.~\ref{fig:fig1}C red curve and Fig.~\ref{fig:fig1}E). Finally, after waiting for heat to diffuse into the PU, a third PAT image showed a slight signal decrease in the PU target and a continued signal increase in the agar target (Fig.~\ref{fig:fig1}C green dashed curve and Fig.~\ref{fig:fig1}F), capturing the different $d\Gamma/dT$ in the two materials and demonstrating $d\Gamma/dT$ contrast. Therefore, in a single imaging cycle, HEPAT generates images of optical absorption, RF absorption, and $d\Gamma/dT$ (Fig.~\ref{fig:fig1}D--F).

\begin{figure}[t]
    \centering
    \includegraphics[width=0.95\linewidth]{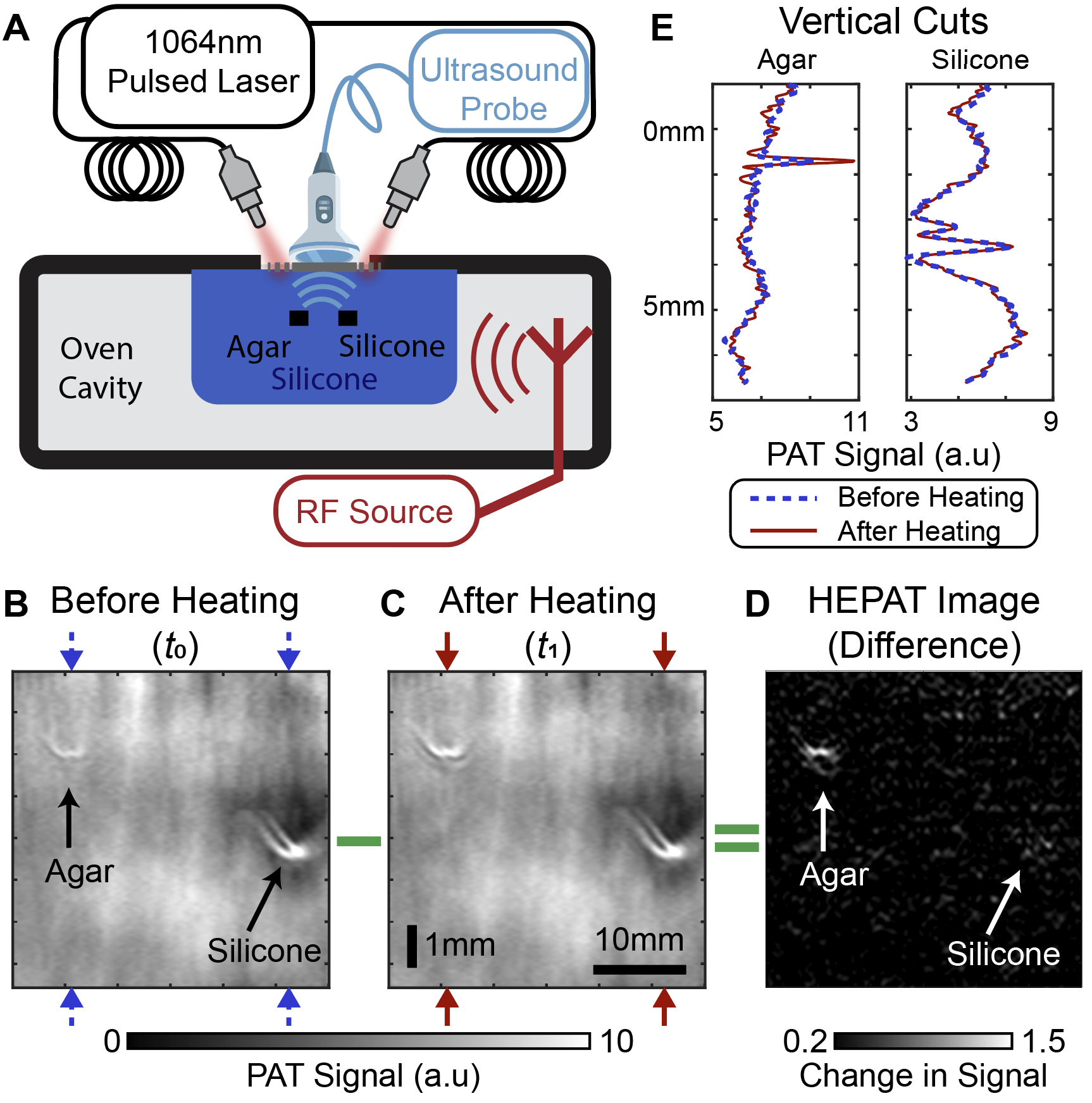}
    \caption{Microwave oven-based HEPAT demonstrating RF absorption contrast with a US\$50 addition to an existing PAT system. (A) Schematic of experimental setup. Agar (left) and silicone (right) targets are placed inside a consumer microwave oven and illuminated by a pulsed laser through an Al foil mesh window. The resulting PA signals are detected by a probe above the targets. (B) PA image of the top surface of the targets before microwave heating. (C) PA image taken after RF heating, which raises the agar's temperature while leaving the silicone largely unaffected. (D) HEPAT image, obtained by subtracting the pre-heating (B) from the post-heating (C) images. Because only the agar heated, only it showed a clear increase in signal, whereas the silicone remained unchanged. See Visualization 1 for the HEPAT image sequence acquired during heating. (E) Vertical cut profiles through the agar and silicone regions confirm that the agar's PA amplitude increases markedly, whereas the silicone signal stays nearly constant.}
    \label{fig:oven-HEPAT}
\end{figure}

To validate HEPAT's ability to image RF absorption using extremely cheap RF systems ($\sim$US\$50), we constructed a tissue phantom and imaged it inside a consumer microwave oven during heating (5~s of 2.45~GHz RF at 1~kW). The phantom consisted of a black-dyed agar target (RF absorber) and a silicone target (minimal RF absorber) embedded in a translucent silicone matrix (Fig.~\ref{fig:oven-HEPAT}A). Silicone, similar to PU, serves as a model for fatty tissue. The targets were placed 25~mm apart to limit thermal diffusion between them, allowing us to image at $t_1$. A 12~ns pulsed 1,064~nm Nd:YAG laser delivered the PA excitation via fiber bundles and slits cut in 10~$\mu$m-thick Al foil. The foil confined the RF to the sample while allowing >80\% of the acoustic signal to pass, because it is $\sim100 \times$ thinner than the wavelength. For detection, we used a commercial ultrasound probe with center frequency of 2.5~MHz and 95\% bandwidth (Jiarui Ultrasound P4-1B). Because only the agar absorbed RF, it heated by $\sim20^\circ$C, exhibiting a marked increase in its PA signal (Fig.~\ref{fig:oven-HEPAT}B--D). In contrast, the silicone target remained at nearly the same temperature and showed negligible signal change. Subtracting the pre- at $t_0$ (Fig.~\ref{fig:oven-HEPAT}B) from the post-heating at $t_1$ PA images (Fig.~\ref{fig:oven-HEPAT}C) reveals the HEPAT image (Fig.~\ref{fig:oven-HEPAT}D), clearly delineating the RF-absorbing agar while the non-absorbing silicone is not visible. This demonstrates that HEPAT can effectively map RF absorption in tissue-like samples without the complex RF subsystems needed for TAT.

\begin{figure*}[t]
    \centering
    \includegraphics[width=0.95\textwidth]{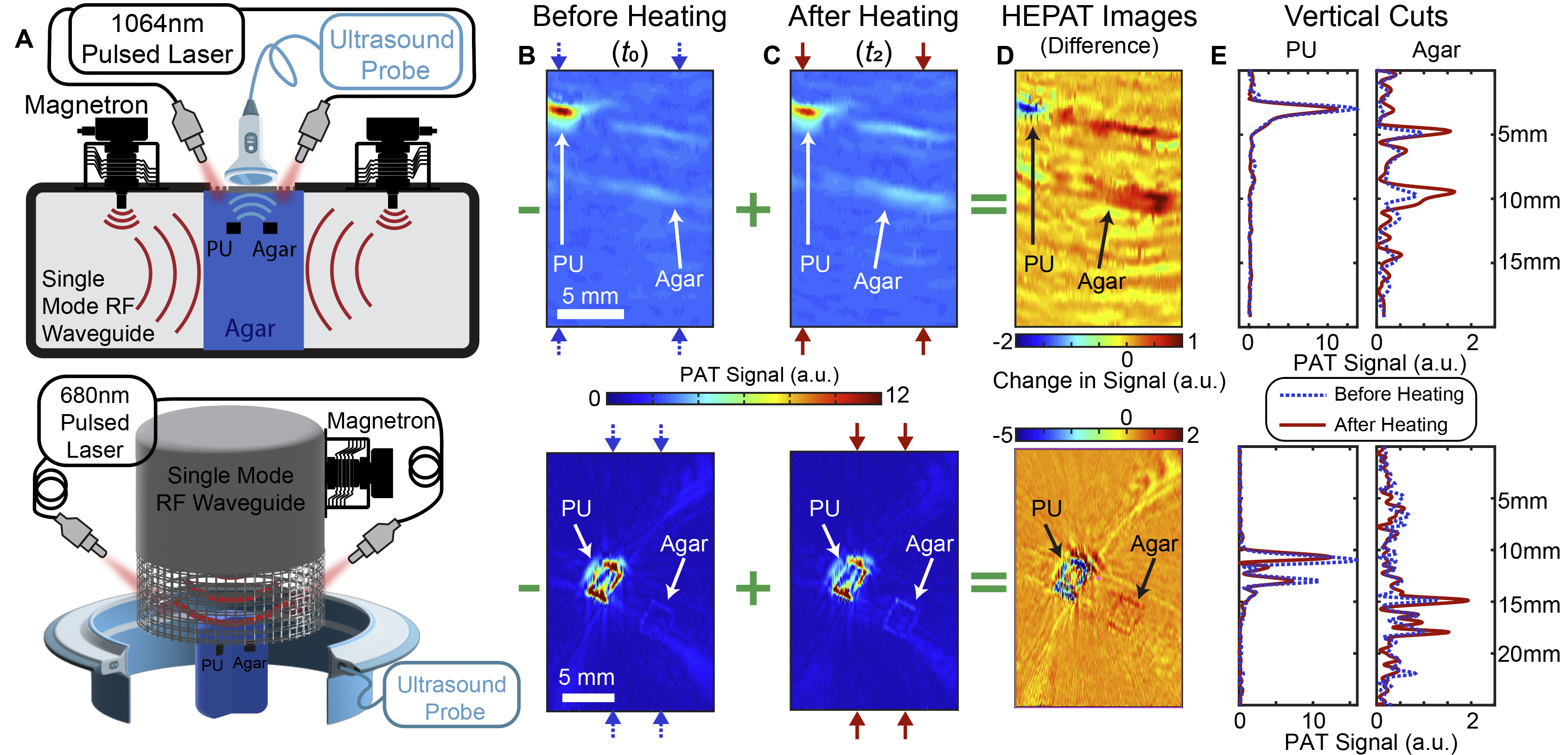}
    \caption[HEPAT on specifically engineered setups demonstrating Gr\"uneisen contrast $d\Gamma/dT$]{HEPAT on specifically engineered setups demonstrating Gr\"uneisen contrast $d\Gamma/dT$. Top row shows the images from the linear probe experiment, and the bottom row shows the images in the ring probe experiment. (A) Drawings of both setups where consumer magnetrons heat fatty (PU) and watery (agar) tissue phantoms and ultrasound probes detect the PA signals. (B, C) Photoacoustic images of each phantom acquired before (B) and after (C) RF heating and thermal diffusion. (D) HEPAT difference images (after minus before), revealing that heating decreases the PA signal in the PU target while increasing it in the agar---an effect driven by their opposite signs of $d\Gamma/dT$. (E) Vertical cross sections through the PU and agar regions confirm these opposing signal changes, illustrating clear contrast between the two materials that emulates RF absorption contrast.}
    \label{fig:linear-ring}
\end{figure*}

We further demonstrated HEPAT with specifically engineered setups and on tissue-like samples composed mostly of water, consisting of an RF-absorbing agar matrix with non-RF-absorbing PU insertions. The insertions were touching the matrix, so when the matrix absorbed RF, the heat quickly diffused to the insertions. The time for heat to diffuse one acoustic wavelength ($\sim 1.5$~mm) into the PU regions was less than the heating time ($\sim 5$~s). Thus, at the end of the heating period, when the post-heating PA image was acquired, the PU volume that generated most of the PA signal was already at an elevated temperature; therefore, the post-heating image corresponded to time $t_2$.

To image these phantoms, we created two PA systems with integrated RF heaters: a linear ultrasound system (Fig.~\ref{fig:linear-ring}A, top) and a system where the ultrasound detector was a ring (Fig.~\ref{fig:linear-ring}A, bottom). The phantoms used in both systems consisted of an agarose block implanted with dyed PU and agar targets. For the linear system, the RF illumination was provided by two 1~kW 2.45~GHz magnetrons and was delivered through two single-mode TE$_{10}$ rectangular waveguides placed on opposite sides of the phantom. For the ring system, an end-fire TE$_{11}$ single mode circular waveguide delivered 1~kW from a single magnetron. Each waveguide was tuned and acted as a resonant transformer for efficient power coupling into the sample. These RF configurations provided quite uniform heating due to the fundamental mode shapes. The optical delivery and ultrasound probe for the linear system were identical to the experiment in the oven (Fig.~\ref{fig:linear-ring}A, top). For the ring system, 680~nm laser light was delivered through fiber bundles placed around the sample and wire mesh at the end of the waveguide allowed laser light to shine through to the phantom yet, contained the RF inside (Fig.~\ref{fig:linear-ring}A, bottom). A custom-made 256-element full-ring ultrasound detector array (Imasonic Inc.; 5.5~MHz central frequency; >60\% -6 dB bandwidth, 5~cm radius, 26~mm height) was used for detection. As with the linear system, a sheet of 10~$\mu$m-thick Al foil protected the ring array from the high-power RF. Both ultrasound systems fed into two customized 128-channel preamplifier modules with 20 dB gain and two 128-channel data acquisition units (Analogic Corp., SonixDAQ).

We tested HEPAT on the two different setups---first using the linear-array system (Fig.~\ref{fig:linear-ring}, top) and then a ring-array system (Fig.~\ref{fig:linear-ring}, bottom). We acquired an initial PAT image at $t_0$ (Fig.~\ref{fig:linear-ring}B), heated the samples for ~5~s ($\sim +10^\circ$C), and then captured a second image at $t_2$ (Fig.~\ref{fig:linear-ring}C). Subtracting the two images to form the HEPAT image (Fig.~\ref{fig:linear-ring}D) showed that the agar's signal increased while the PU's signal decreased (see cut plots Fig.~\ref{fig:linear-ring}E)---a consequence of their opposite $d\Gamma/dT$. Although the PU did not absorb RF, the surrounding agar did and conducted heat into it, warming its surface enough to reduce its PA amplitude. This led to the stark "cool-blue" PU region versus the "hot-red" agar region in the HEPAT image (Fig.~\ref{fig:linear-ring}D), clearly highlighting the difference between them. In both experiments, the PAT images largely show the amount of dye in each target, whereas only the HEPAT images conclusively reveal that the two targets were made of different materials.

Although thermal diffusion prevents us from isolating $\mu_{RF}$ in Fig.~\ref{fig:linear-ring}, we can still discriminate between the phantoms via $d\Gamma/dT$, a contrast that is physically distinct from $\mu_{RF}$ contrast, yet uniquely mapped to it: materials with low $\mu_{RF}$ have negative $d\Gamma/dT$ (fatty tissue/phantoms), and materials with high $\mu_{RF}$ have positive $d\Gamma/dT$ (watery tissue/phantoms). Thus, even though $t_1$ and $\mu_{RF}$ are not available, $d\Gamma/dT$ contrast functionally replaces $\mu_{RF}$.

In sum, we propose and experimentally validate with phantoms a new multi-modality imaging method---HEPAT---that extends the capabilities of PAT with minimal added cost. By integrating a low-cost RF heating system---composed of household microwave oven components---we significantly boost image contrast. We show that HEPAT exploits both available endogenous types of contrast (electromagnetic and thermomechanical), can map three material properties in a single imaging cycle, and highlights fatty and watery samples clearly.

We have shown how adjusting the imaging sequence can isolate specific material parameters (e.g., $t_1$ maps $\mu_{\mathrm{RF}}$ and $t_2$ maps $d\Gamma/dT$).
HEPAT can be extended to reveal additional contrasts, but these contrasts can be confounded and difficult to isolate.
For example, temperature-induced changes in sound speed enable mapping of $dc/dT$ by tracking spatial shifts; however, this same sound-speed heterogeneity generates registration artifacts in pre- and post-heating images that require image-registration algorithms.
Here, we low-pass filtered, manually shifted, and used different values of $c$ in the delay-and-sum reconstruction to register images properly, but $dc/dT$ heterogeneity still caused feature misalignment.
Further, heat transport may be mapped by tracking features over time; however, uncertainties in heat transport preclude absolute $d\Gamma/dT$ recovery: without direct temperature measurements or modeling, only the sign of $d\Gamma/dT$ can be determined.

For in vivo use, the $10\,^{\circ}\mathrm{C}$ heating demonstrated in the ring system is expected to correspond to a safe thermal dose under the CEM43 framework \cite{yarmolenko2011thresholds}: a conservative COMSOL bioheat simulation of an in vivo configuration predicts a peak thermal dose of 3~min CEM43.  
Future studies can further reduce temperatures by using periodic RF heating with lock-in detection to improve SNR \cite{tzang2016lockin}; PAT has detected temperature changes as small as $0.2^\circ$C via repeated measurements and averaging \cite{pramanik2009thermoacoustic}. 
Our initial tests on ex vivo samples and live mice suggest that improved feature alignment and PA contrast are still needed for robust in vivo imaging. 
Nonetheless, the phantom results here establish the feasibility of HEPAT and its ability to enhance contrast, motivating future studies in biological tissues.

Last, there is an opportunity to develop a new family of RF-switchable probes. Periodic RF heating would allow for lock-in detection, enabling sensitive deep-tissue imaging \cite{feng2014tmpi}. These agents will not suffer from photobleaching and may be switched at greater depths than visible-wavelength probes.

\begin{backmatter}

    \bmsection{Funding}
    National Natural Science Foundation of China (62475129, 61735016 and 61871251); Youth Innovation Fund of Beijing National Research Center for Information Science and Technology; Global Engineering Internship Program, Stanford University; Beijing Natural Science Foundation (L256003); Xishan–Tsinghua Industry-Academia-Research Integration Initiative; Tsinghua University-Fuzhou Institute of Data Technology; STI2030-Major Projects (2022ZD0212000).

    \bmsection{Acknowledgment}
    We thank Manxiu Cui and Hongzhi Zuo for useful discussions.

    \bmsection{Disclosures}
    The authors declare no conflicts of interest.

    \bmsection{Data Availability}
    Data underlying the results presented in this paper are not publicly available at this time but may be obtained from the authors upon reasonable request.

\end{backmatter}

\bibliography{references}

@article{wang2009multiscale,
  author = {Wang, L. V.},
  title = {Multiscale photoacoustic microscopy and computed tomography},
  journal = {Nat. Photonics},
  volume = {3},
  pages = {503--509},
  year = {2009}
}

@article{fu2019photoacoustic,
  author = {Fu, Q. and Zhu, R. and Song, J. and others},
  title = {Photoacoustic imaging: Contrast agents and their biomedical applications},
  journal = {Adv. Mater.},
  volume = {31},
  pages = {1805875},
  year = {2019}
}

@article{yang2009nanoparticles,
  author = {Yang, X. and Stein, E. W. and Ashkenazi, S. and others},
  title = {Nanoparticles for photoacoustic imaging},
  journal = {Wiley Interdiscip. Rev. Nanomed. Nanobiotechnol.},
  volume = {1},
  pages = {360--368},
  year = {2009}
}

@article{cox2009estimating,
  author = {Cox, B. T. and Arridge, S. R. and Beard, P. C.},
  title = {Estimating chromophore distributions from multiwavelength photoacoustic images},
  journal = {J. Opt. Soc. Am. A},
  volume = {26},
  pages = {443--455},
  year = {2009}
}

@article{ku2005thermoacoustic,
  author = {Ku, G. and Fornage, B. D. and Jin, M. and Xu, X. and Hunt, K. and Wang, L. V.},
  title = {Thermoacoustic and photoacoustic tomography of thick biological tissues toward breast imaging},
  journal = {Technol. Cancer Res. Treat.},
  volume = {4},
  pages = {559--565},
  year = {2005}
}

@article{ke2012performance,
  author = {Ke, H. and Erpelding, T. N. and Jankovic, L. and Liu, C. and Liu, L. and Wang, L. V.},
  title = {Performance characterization of an integrated ultrasound, photoacoustic, and thermoacoustic imaging system},
  journal = {J. Biomed. Opt.},
  volume = {17},
  pages = {056010},
  year = {2012}
}

@article{gabriel1996dielectric,
  author = {Gabriel, S. and Lau, R. W. and Gabriel, C.},
  title = {The dielectric properties of biological tissues: III. Parametric models for the dielectric spectrum of tissues},
  journal = {Phys. Med. Biol.},
  volume = {41},
  pages = {2271--2293},
  year = {1996}
}

@article{lazebnik2007large,
  author = {Lazebnik, M. and Popovic, D. and McCartney, L. and others},
  title = {A large-scale study of the ultrawideband microwave dielectric properties of normal, benign and malignant breast tissues obtained from cancer surgeries},
  journal = {Phys. Med. Biol.},
  volume = {52},
  pages = {6093--6115},
  year = {2007}
}

@article{pramanik2009thermoacoustic,
  author = {Pramanik, M. and Wang, L. V.},
  title = {Thermoacoustic and photoacoustic sensing of temperature},
  journal = {J. Biomed. Opt.},
  volume = {14},
  pages = {054024},
  year = {2009}
}

@article{shah2008photoacoustic,
  author = {Shah, J. and Park, S. and Aglyamov, S. and others},
  title = {Photoacoustic imaging and temperature measurement for photothermal cancer therapy},
  journal = {J. Biomed. Opt.},
  volume = {13},
  pages = {034024},
  year = {2008}
}

@article{larina2005real,
  author = {Larina, I. V. and Larin, K. V. and Esenaliev, R. O.},
  title = {Real-time optoacoustic monitoring of temperature in tissues},
  journal = {J. Phys. D: Appl. Phys.},
  volume = {38},
  pages = {2633--2639},
  year = {2005}
}

@article{nikitin2012temperature,
  author = {Nikitin, S. M. and Khokhlova, T. D. and Pelivanov, I. M.},
  title = {Temperature dependence of the optoacoustic transformation efficiency in ex vivo tissues for application in monitoring thermal therapies},
  journal = {J. Biomed. Opt.},
  volume = {17},
  pages = {061214},
  year = {2012}
}

@article{lou2010temperature,
  author = {Lou, C. and Xing, D.},
  title = {Temperature monitoring utilising thermoacoustic signals during pulsed microwave thermotherapy: A feasibility study},
  journal = {Int. J. Hyperthermia},
  volume = {26},
  number = {4},
  pages = {338--346},
  year = {2010}
}

@article{Wang2011ThermalIVPA,
  author    = {Bo Wang and Stanislav Emelianov},
  title     = {Thermal intravascular photoacoustic imaging},
  journal   = {Biomedical Optics Express},
  year      = {2011},
  volume    = {2},
  number    = {11},
  pages     = {3072--3078},
  doi       = {10.1364/BOE.2.003072},
  pmid      = {22076268},
  pmcid     = {PMC3207376}
}

@article{Shi2019,
  author  = {Shi, Junhui and Wong, Terence T. W. and He, Yun and Li, Lei and Zhang, Ruiying and Yung, Christopher S. and Hwang, Jeeseong and Maslov, Konstantin and Wang, Lihong V.},
  title   = {High-resolution, high-contrast mid-infrared imaging of fresh biological samples with ultraviolet-localized photoacoustic microscopy},
  journal = {Nature Photonics},
  volume  = {13},
  number  = {9},
  pages   = {609--615},
  year    = {2019},
  month   = sep,
  issn    = {1749-4893},
  doi     = {10.1038/s41566-019-0441-3},
  url     = {https://doi.org/10.1038/s41566-019-0441-3}
}

@article{Ma2016Grueneisen,
  author  = {Ma, Jun and Shi, Junhui and Hai, Pengfei and Zhou, Yong and Wang, Lihong V.},
  title   = {Grueneisen relaxation photoacoustic microscopy in vivo},
  journal = {Journal of Biomedical Optics},
  volume  = {21},
  number  = {6},
  pages   = {066005},
  year    = {2016},
  doi     = {10.1117/1.JBO.21.6.066005}
}

@article{Liang2018GrueneisenLipids,
  author  = {Liang, Simon and Lashkari, Bahman and Choi, Sung Soo Sean and Ntziachristos, Vasilis and Mandelis, Andreas},
  title   = {The application of frequency-domain photoacoustics to temperature-dependent measurements of the Gr{\"u}neisen parameter in lipids},
  journal = {Photoacoustics},
  volume  = {11},
  pages   = {56--64},
  year    = {2018},
  issn    = {2213-5979},
  doi     = {10.1016/j.pacs.2018.07.005}
}

@article{pramanik2008breast,
  author  = {Pramanik, Manojit and Ku, Geng and Li, Changhui and Wang, Lihong V.},
  title   = {Design and evaluation of a novel breast cancer detection system combining both thermoacoustic ({TA}) and photoacoustic ({PA}) tomography},
  journal = {Med. Phys.},
  year    = {2008},
  volume  = {35},
  number  = {6},
  pages   = {2218--2223},
  doi     = {10.1118/1.2911157}
}

@article{tian2015dualpulse,
  author  = {Tian, Chao and Xie, Zhixing and Fabiilli, Mario L. and Wang, Xueding},
  title   = {Imaging and sensing based on dual-pulse nonlinear photoacoustic contrast: a preliminary study on fatty liver},
  journal = {Opt. Lett.},
  year    = {2015},
  volume  = {40},
  number  = {10},
  pages   = {2253--2256},
  doi     = {10.1364/OL.40.002253}
}

@article{yarmolenko2011thresholds,
  author  = {Yarmolenko, P. S. and Moon, E. J. and Landon, C. and Manzoor, A. and Hochman, D. W. and Viglianti, B. L. and Dewhirst, M. W.},
  title   = {Thresholds for thermal damage to normal tissues: An update},
  journal = {Int. J. Hyperthermia},
  year    = {2011},
  volume  = {27},
  number  = {4},
  pages   = {320--343},
  doi     = {10.3109/02656736.2010.534527}
}

@article{tzang2016lockin,
  author  = {Tzang, O. and Piestun, R.},
  title   = {Lock-in detection of photoacoustic feedback signal for focusing through scattering media using wave-front shaping},
  journal = {Opt. Express},
  year    = {2016},
  volume  = {24},
  number  = {24},
  pages   = {28122--28130},
  doi     = {10.1364/OE.24.028122}
}

@article{feng2014tmpi,
  author  = {Feng, Xiaohua and Gao, Fei and Zheng, Yuanjin},
  title   = {Thermally modulated photoacoustic imaging with super-paramagnetic iron oxide nanoparticles},
  journal = {Opt. Lett.},
  year    = {2014},
  volume  = {39},
  number  = {12},
  pages   = {3414--3417},
  doi     = {10.1364/OL.39.003414}
}


\end{document}